\documentstyle[preprint,pre,aps,amstex,epsfig]{revtex}
\begin{document}
\draft
\title{Diffusion effects on the breakdown of a linear amplifier model
driven by the square of a Gaussian field}
\author{A. Asselah$^a$, P. Dai Pra$^b$, J. L. Lebowitz$^c$, and Ph.
Mounaix$^d$}
\address{$^a$LATP, UMR 6632 du CNRS, Centre de Math\'ematique et
Informatique, Universit\'e de Provence, 39 rue F. Joliot-Curie, 13453
Marseille Cedex 13, France.}
\address{$^b$Dipartimento di Matematica, Politecnico di Milano
Piazza Leonardo da Vinci 32, I-20133 Milano, Italy.}
\address{$^c$Departments of Mathematics and Physics, Rutgers, The State
University of New Jersey, Piscataway, New Jersey 08854-8019.}
\address{$^d$Centre de Physique Th\'eorique, UMR 7644 du CNRS, Ecole
Polytechnique, 91128 Palaiseau Cedex, France.}
\date{\today}
\maketitle
\begin{abstract}
We investigate solutions to the equation $\partial_t{\cal E} - {\cal
D}\Delta {\cal E} = \lambda S^2{\cal E}$, where $S(x,t)$ is a Gaussian
stochastic field with covariance $C(x-x',t,t')$, and $x\in {\mathbb
R}^d$. It is shown that the coupling $\lambda_{cN}(t)$ at which the
$N$-th moment $\langle {\cal E}^N(x,t)\rangle$ diverges at time $t$, is
always less or equal for ${\cal D}>0$ than for ${\cal D}=0$. Equality
holds under some reasonable assumptions on $C$ and, in this case,
$\lambda_{cN}(t)=N\lambda_c(t)$ where $\lambda_c(t)$ is the value of
$\lambda$ at which $\langle\exp\lbrack
\lambda\int_0^tS^2(0,s)ds\rbrack\rangle$ diverges. The ${\cal D}=0$
case is solved for a class of $S$. The dependence of $\lambda_{cN}(t)$
on $d$ is analyzed. Similar behavior is conjectured when diffusion is
replaced by diffraction, ${\cal D}\rightarrow i{\cal D}$, the case of
interest for backscattering instabilities in laser-plasma interaction.
\end{abstract}
\section{Introduction}\label{sec1}
We investigate the development of a linear amplification in a system
driven by the square of a Gaussian noise. This problem arose and
continues to be of interest in modeling the backscattering of an
incoherent high intensity laser light by a plasma. There is a large
litterature on this topic, and we refer the interested reader to Ref.\
\cite{Laval} for background. Our starting point here is the work by
Rose and Dubois\ \cite{RD} who investigated the following equation for
the complex amplitude ${\cal E} (x,z)$ of the scattered electric field
\begin{equation}\label{eq1.1}
\left\lbrace\begin{array}{l}
\partial_z{\cal E}(x,z)-i{\cal D}\Delta
{\cal E}(x,z)=
\lambda\vert S(x,z)\vert^2{\cal E}(x,z),\\
z\in [0,L],\ x\in\Lambda\subset {\mathbb R}^2,\ {\rm and}\
{\cal E}(x,0)={\cal E}_0(x).
\end{array}\right.
\end{equation}
In Eq.\ (\ref{eq1.1}), $z$ and $x$ correspond to the axial and
transverse directions in a plasma of length $L$ and cross-sectional
domain $\Lambda$. The input at $z=0$, ${\cal E}_0(x)$, is a given
function of $x$ and $\Lambda$ will be generally taken to be a torus
(e.g. in numerical solutions of Eq.\ (\ref{eq1.1}) using spectral
methods). The coupling constant $\lambda >0$ is proportional to the
average laser intensity and ${\cal D}$ is a constant parameter
introduced for convenience. The complex amplitude of the laser electric
field $S(x,z)$ is a homogeneous Gaussian stochastic field defined by
\begin{eqnarray*}
&&\langle S(x,z)\rangle =\langle S(x,z)S(x',z')\rangle =0,\\
&&\langle S(x,z)S(x',z')^\ast\rangle =C(x-x',z-z'),
\end{eqnarray*}
where the correlation function $C(x,z)$ is the solution to
\begin{equation}\label{eq1.2}
\left\lbrace\begin{array}{l}
\partial_zC(x,z)+\frac{i}{2}\Delta C(x,z)=0,\\
z\in [0,L],\ x\in\Lambda,\ {\rm and}\
C(x,0)={\cal C}(x),
\end{array}\right.
\end{equation}
with ${\cal C}(x)$ a given function of $x$\ \cite{note1}, normalized
so that ${\cal C}(0)\equiv\langle\vert S(x,z)\vert^2\rangle=1$.
\paragraph*{}Using heuristic arguments and numerical simulations, Rose
and DuBois found that the expected value of the energy density of the
scattered field $\langle\vert {\cal E}(x,L)\vert^2\rangle$ diverged for
every $L>0$ as $\lambda$ increased to some critical value
$\lambda_c(L)$. The average $\langle\vert {\cal E}\vert^2\rangle$ is
over the realizations of the Gaussian field $S$. This divergence
indicates a breakdown in the assumptions made in deriving Eq.\
(\ref{eq1.1}), which neglects both nonlinear saturation and transient
time evolution\ \cite{RD,PRE95}. Physically, it can be interpreted as
indicating a change in the nature of the amplification caused by the
plasma.
\paragraph*{}To see the origin of this divergence in its simplest form,
consider the case where ${\cal D}$ is set equal to zero in Eq.\
(\ref{eq1.1}), and neglect all dependence of $S$ on $x$ and $z$. We are
then led to the equation
\begin{equation}\label{eq1.3}
\frac{d{\cal E}(z)}{dz} = \lambda S^2 {\cal E}(z),
\end{equation}
which yields
$$
{\cal E}(z) = {\cal E}(0) e^{\lambda S^2 z}, \quad z>0.
$$
Here $S^2 =S^2_1 +S^2_2$ and $S_1$, $S_2$ are two independent real
Gaussian random variables with zero mean and unit variance. It is easily
seen that the probability distribution of ${\cal E}(z)$, setting ${\cal
E}(0)=1$, has the density
\begin{equation}\label{eq1.4}
W({\cal E},z) = (2\lambda z)^{-1} {\cal E}^{-[1+(2\lambda z)^{-1}]}
\quad {\rm for} \quad  {\cal E} \geq 1, \quad z>0.
\end{equation}
If we now take moments of ${\cal E}$ at some value $L$ of $z$, we find
that $\langle {\cal E}^N(L) \rangle$ will diverge whenever $2N\lambda
L\geq 1$. At the critical coupling $\lambda_{cN}(L)=(2NL)^{-1}$, there
is a qualitative transition of the amplification of $\langle {\cal
E}^N(L) \rangle$ from a regime where it is dominated by the bulk of the
order-one-fluctuations of $S$ to a regime where it is dominated by the
large fluctuations of $S$. This toy model can be thought
of as an idealization in which the size of the plasma is very small
compared to the correlation length of the laser field. This is
certainly not a reasonable physical approximation and we shall later
consider situations in which $S$ in Eq.\ (\ref{eq1.3}) is $z$-dependent
with a covariance $C(z,z')$.  The equation is then still solvable more
or less explicitly, depending on the form of $C$, at least as far as
the dependence of the divergence of the moments of ${\cal E}$ on
$\lambda$ and $L$ is concerned. The main difference from Eq.\
(\ref{eq1.4}) is that for small enough values of $\lambda$, the first
few moments need not diverge for any $L$.
\paragraph*{}In this paper, we extend these results to the $x$-dependent
case where $i{\cal D}$ in Eq.\ (\ref{eq1.1}) is replaced by ${\cal D}$,
i.e.\ we consider a diffusive process in $x$ rather than a diffractive
one. Somewhat surprisingly the diffusion does not suppress the onset of
divergences in moments of the field. This suggests a similar behavior
for the diffractive case -- in accord with the numerical results of\
\cite{RD} -- but we are unable to prove this at the present time.
\paragraph*{}Before going on to the formulation and presentation of
results for the diffusive case, we make some remarks about the relation
between expectations over different realizations of the Gaussian
driving term $|S|^2$ and the outcome of a given experiment. Accepting
the idealizations inherent in assuming Gaussian statistics and neglect
of nonlinear terms, the physically relevant question relating to the
solution of the stochastic PDE\ (\ref{eq1.1}) appears to be the
following: What is the probability that for given $\Lambda$ and $L$
there will be small regions in $\Lambda$ through which a significant
fraction of the total incoming power is backscattered, (here "total"
means through the whole domain $\Lambda$). Put more physically, imagine
$\Lambda$ to be divided up into $M\gg 1$ cells of equal area
$\vert\Lambda\vert /M$ and let $R\gg 1/M$ be a specified number. We
want to compute the probability $P$ that in at least one of the cells
the integral of $|{\cal E}|^2$ over that cell exceeds
$R\vert\Lambda\vert$. In the case where ${\cal D}$ is set equal to
zero, this can be answered by taking for the cell size the transverse
correlation length of $\vert S\vert^2$ and assuming the field inside
each cell to be transversally constant and evolving along $z$ under
Eq.\ (\ref{eq1.3}) with a $z$-dependent $S$. One finds that $P$ greatly
increases as $\lambda$ passes its critical value for the divergence of
the second moment, from $P\ll 1$ for $\lambda <\lambda_{c2}(L)$ to
$P\simeq 1$ for $\lambda >\lambda_{c2}(L)$. We expect that this
probability will behave similarly in real systems.
\paragraph*{}The outline of the rest of this paper is as follows. In
Sec.\ \ref{sec2} we introduce our diffusion-amplification model. In
Sec.\ \ref{sec3} we prove that the value of the critical coupling
obtained without the diffusion term cannot be less than the one
obtained with the diffusion term. In Sec.\ \ref{sec4} we prove that for
a large class of Gaussian fields $S$ the values of the critical
coupling obtained with or without the diffusion term are the same.
Section\ \ref{sec5} is devoted to the explicit solution of the
diffusion-free problem in the particular case where the on-axis field
$S(0,z)$ is a linear functional of a Gauss-Markov process. {}Finally, in
Sec.\ \ref{sec6} we study the dependence of the critical coupling on
the space dimensionality in the case of a factorable correlation
function $C$.
%
%
\section{Model and definitions}
\label{sec2}
As explained in the introduction, we consider a modified version of the
linear convective amplifier model obtained by replacing $i{\cal D}$ by
${\cal D}$ on the left-hand side of Eq.\ (\ref{eq1.1}). Taking ${\cal
D}=1/2$ without loss of generality, one is thus led to the problem
\begin{equation}\label{eq2.1}
\left\lbrace\begin{array}{l}
\partial_t{\cal E}(x,t)-\frac{1}{2}\Delta
{\cal E}(x,t)=
\lambda S(x,t)^2{\cal E}(x,t),\\
t\in [0,T],\ x\in {\mathbb R}^d,\ {\rm and}\
{\cal E}(x,0)={\cal E}_0(x),
\end{array}\right.
\end{equation}
where, following the usual notation used in diffusion problems, the
time variable $t$ (resp. $T$) plays the role of the axial space variable
$z$ (resp. $L$). In Eq.\ (\ref{eq2.1}), we restrict ourselves to the
cases where $S(x,t)$ is a real homogeneous Gaussian field defined by
\begin{eqnarray*}
&&\langle S(x,t)\rangle =0,\\
&&\langle S(x,t)S(x',t')\rangle =C(x-x',t,t'),
\end{eqnarray*}
with the normalization $C(0,0,0)\equiv\langle S(x,0)^2\rangle=1$, and
we will take ${\cal E}_0(x)\equiv 1$ as an initial condition. Note that
$S(x,t)$ is not assumed to be stationary in $t$, and that the rest of our
analysis is essentially unaffected if we replace ${\mathbb R}^d$ by a
$d$-dimensional torus.
\paragraph*{}The critical coupling $\lambda_{cN}(T)$ and its
diffusion-free counterpart $\overline{\lambda}_{cN}(T)$ are defined by
\begin{mathletters}\label{eq2.2}
\begin{eqnarray}
&&\lambda_{cN}(T)=
\inf\lbrace \lambda>0:\langle {\cal E}(0,T)^N\rangle =+\infty\rbrace,
\label{eq2.2a}\\
&&\overline{\lambda}_{cN}(T)=
\inf\lbrace \lambda>0:\langle {\rm e}^{N\lambda\int_0^T
S(0,t)^2dt}\rangle =+\infty\rbrace,
\label{eq2.2b}
\end{eqnarray}
\end{mathletters}
where $\langle .\rangle$ denotes the average over the realizations of
$S$. {}For a given $T>0$, Eqs.\ (\ref{eq2.2}) gives the value of $\lambda$
at which $\langle {\cal E}(x,T)^N\rangle$ blows up with and without
diffusion respectively.
\paragraph*{}Finally, in order not to make the calculations too
cumbersome, we will use in the following the compact notation
\begin{eqnarray*}
&&{\bf t}\equiv (n,t),\\
&&\int d{\bf t}\equiv \sum_{n=1}^N\int_0^Tdt,\\
&&S({\bf t})\equiv S(x_n(t),t),\\
&&C({\bf s},{\bf t})\equiv\langle S({\bf
s})S({\bf t})\rangle =C(x_n(s)-x_m(t),s,t),\\
&&C_0({\bf s},{\bf t})\equiv C(0,s,t),\\
&&(\varphi , \psi) = \int
\varphi({\bf t}) \psi({\bf t}) d{\bf t},
\end{eqnarray*}
with $s,t\in [0,T]$, $n,m\in{\mathbb N} $ ($1\le n,m\le N$), and where
the $x_n(\cdot)$ are given continuous paths on ${\mathbb R}^d$. The
covariance operators $\hat{T}_C$ and $\hat{t}_{C_0}$, respectively
acting on $\varphi({\bf t})\in L^2(d{\bf t})$ and $\varphi(t)\in
L^2(dt)$, are defined by
\begin{eqnarray*}
&&(\hat{T}_C\varphi)({\bf s})=
\int C({\bf s},{\bf t})\varphi({\bf t})\, d{\bf t},\\
&&(\hat{t}_{C_0}\varphi)(s)=
\int_0^T C(0,s,t)\varphi(t)\, dt.
\end{eqnarray*}
%
%
\section{Comparison of $\lambda_{cN}(T)$ and $\overline{\lambda}_{cN}(T)$}
\label{sec3}
In this section we prove that
$\lambda_{cN}(T)\le\overline{\lambda}_{cN}(T)$. We begin with two
technical lemmas that will be useful in the following.
\paragraph*{}
\bigskip
\noindent {\bf Lemma 1}:
Suppose the covariance function $C(x,t,t')$ is continuous. Let
$\mu_1^{x({\bf t})}\ge \mu_2^{x({\bf t})}\ge ...\ge 0$ be the
eigenvalues of the covariance operator $\hat{T}_C$. Here, the
superscript $x({\bf t})$ denotes the $N$ continuous paths $x_n(t)$,
$1\le n\le N$. Then $\langle\exp\lambda\int S({\bf t})^2d{\bf t}\rangle
<+\infty$ if and only if $\lambda <(2\mu_1^{x({\bf t})})^{-1}$, and in
this case one has
\begin{equation}\label{eq3.1}
\log\langle {\rm e}^{\lambda\int S({\bf
t})^2d{\bf t}}\rangle =-\frac{1}{2}\sum_{i\ge 1}\log\left(
1-2\lambda\mu_i^{x({\bf t})}\right) \le \frac{N\lambda
\int_0^TC(0,t,t)\, dt}{1-2\lambda\mu_1^{x({\bf t})}}.
\end{equation}
\paragraph*{}
\bigskip
To show\ (\ref{eq3.1}), consider the Hilbert space of the $L^2(d{\bf
t})$ functions $\varphi(n,t)\equiv \varphi({\bf t})$ with the scalar
product $(\varphi , \psi)$. Since $C({\bf s},{\bf t})$ is continuous in
$({\bf s},{\bf t})$, and therefore bounded in compact sets, we have
that $\int \int C({\bf s},{\bf t})^2 d{\bf s}d{\bf t} < +\infty$. By
\cite{resi}, Theorem VI.23, it follows that the covariance operator is
compact (and self-adjoint) in $L^2(d{\bf t})$. Therefore there is an
orthonormal basis $\{\varphi_j \}_{j \ge 1}$ such that $\hat{T}_C
\varphi_j= \mu_j^{x({\bf t})} \varphi_j$. Consider now the sequence of
random variables $X_j = (S,\varphi_j)$. As linear functionals of the
Gaussian field $S$, the $X_j$'s form a Gaussian sequence with $\langle
X_i\rangle =0$ and $\langle X_i X_j\rangle =(\varphi_i, \hat{T}_C
\varphi_j) =\mu_j^{x({\bf t})} \delta_{ij}$. The equality in Eq.\
(\ref{eq3.1}) is then obtained straightforwardly from $\int S^2({\bf
t})d{\bf t} = \sum_{j=1}^{+\infty} X_j^2$ and the simple Gaussian
identity $\left\langle e^{\lambda X_i^2}\right\rangle =
\left(1-2\lambda\mu_i^{x({\bf t})}\right)^{-1/2}$, for
$2\lambda\mu_i^{x({\bf t})} <1$. The inequality in Eq.\ (\ref{eq3.1})
follows from $-\log (1-x)\le x/(1-x)$ and the fact that
$\sum_i\mu_i^{x({\bf t})}=\int C({\bf t},{\bf t})\, d{\bf t}
=N\int_0^TC(0,t,t)\, dt$.
%
%
\paragraph*{}In the following subsection, $\varphi({\bf t})\equiv
\varphi(n,t)$ will denote a set of $N$ test functions normalized such
that $(\varphi , \varphi) =\sum_{n=1}^N\int_0^T\varphi(n,t)^2\, dt=1$.
\paragraph*{}
\bigskip
\noindent {\bf Lemma 2}:
Assume that for every $T>0$ one has $\lim_{x\rightarrow 0}\sup_{s,t\in
[0,T]} \vert C(x,s,t)-C(0,s,t)\vert =0$. Then, $\forall\varepsilon >0$,
$\exists\delta >0$ such that
$$
\left\vert
(\varphi,\hat{T}_C\varphi)-(\varphi,\hat{T}_{C_0}\varphi)
\right\vert
<\varepsilon
$$
for every $x_n(\cdot)\in B_{\delta ,T}$, $1\le n\le N$, where
$B_{\delta ,T}$ is the set of continuous paths
$x(\cdot)$ such that $\vert x(t)\vert <\delta$ for every $t\in [0,T]$.
\paragraph*{}
\bigskip
The proof of this lemma is straightforward: from the
uniform convergence condition on $C(x,s,t)$ it follows that
$\forall\varepsilon >0$, $\exists\delta >0$ such that $\vert C({\bf
s},{\bf t})-C_0({\bf s},{\bf t})\vert < \varepsilon$ for every
$x_n(\cdot)\in B_{\delta,T}$, $1\le n\le N$. Thus,
$\forall\varepsilon^\prime >0$, $\exists\delta >0$ such that
\begin{eqnarray*}
&&\left\vert
(\varphi,\hat{T}_C\varphi)-(\varphi,\hat{T}_{C_0}\varphi)
\right\vert\le
(\vert\varphi\vert,\hat{T}_{\vert C-C_0\vert}\vert\varphi\vert)\\
&&< \varepsilon^\prime \left(\int\vert\varphi({\bf s})\vert\, d{\bf
s}\right)^2 \le \varepsilon^\prime NT,
\end{eqnarray*}
for every $x_n(\cdot)\in B_{\delta,T}$, $1\le n\le N$. It remains to take
$\varepsilon^\prime =\varepsilon/(NT)$, which proves Lemma 2.
\paragraph*{}We can now state the main result of this section. Namely,
that one of the diffusion effects on the divergence of the moments of
${\cal E}(x,T)$ is a lowering (or, more exactly, a non-increasing) of
the critical coupling. The rigorous formulation of this result can be
stated as the following proposition.
\paragraph*{}
\bigskip
\noindent {\bf Proposition 1}:
{}For every $T>0$, if $\lim_{x\rightarrow 0}\sup_{s,t\in [0,T]} \vert
C(x,s,t)-C(0,s,t)\vert =0$, then $\lambda_{cN}(T) \le
\overline{\lambda}_{cN}(T)$.
\paragraph*{}
\bigskip
In order to prove this proposition, one writes the
moments of ${\cal E}$ in terms of the {}Feynman-Kac formula
\begin{equation}\label{eq3.2}
\langle {\cal E}(0,T)^N\rangle =\left\langle\left\langle
\exp\left\lbrack\lambda
\int S({\bf t})^2\, d{\bf t}\right\rbrack
\right\rangle\right\rangle_{x({\bf t})},
\end{equation}
where $\langle\cdot\rangle_{x({\bf t})}$ denotes a $N$-fold Wiener
integral over $N$ Brownian paths $x_n(t)$, $1\le n\le N$, each arriving
at $x=0$. Let $\lambda > \overline{\lambda}_{cN}(T)$, i.e. $\mu_1
>(2N\lambda)^{-1}$, where $\mu_1$ is the largest eigenvalue of the
covariance operator $\hat{t}_{C_0}$. Let $\phi_1(t)$ be the normalized
eigenfunction associated with $\mu_1$, and $\phi({\bf
t})\equiv\phi(n,t)=N^{-1/2}\phi_1(t)$ for every $1\le n\le N$. [N.B. :
the factor $N^{-1/2}$ ensures the normalization $(\phi,\phi)=1$]. By
definition of $\mu_1^{x({\bf t})}$, one has
\begin{equation}\label{eq3.3}
\mu_1^{x({\bf t})}\ge (\phi,\hat{T}_C\phi).
\end{equation}
By Lemma 2, $\forall\varepsilon >0$, $\exists\delta >0$ such that
\begin{equation}\label{eq3.4}
(\phi,\hat{T}_C\phi)\ge (\phi,\hat{T}_{C_0}\phi)-\varepsilon
=N\mu_1-\varepsilon
\end{equation}
for every $x_n(\cdot)\in B_{\delta ,T}$, $1\le n\le N$. If one now takes
$\varepsilon <N\mu_1-\frac{1}{2\lambda}$, it follows from Eqs.\
(\ref{eq3.3}) and\ (\ref{eq3.4}) that $\mu_1^{x({\bf
t})}>1/2\lambda$ and so, by Lemma 1,
$$
\left\langle\exp\left\lbrack\lambda\int S({\bf t})^2\, d{\bf
t}\right\rbrack\right\rangle =+\infty
$$
for every $x_n(\cdot)\in B_{\delta ,T}$, $1\le n\le N$. {}Finally, since
the set of the Brownian paths $x_n(t)$ that are in $B_{\delta ,T}$ has
a strictly positive Wiener measure, one finds from Eq.\ (\ref{eq3.2})
that $\langle {\cal E}(0,T)^N\rangle =+\infty$, so $\lambda\ge
\lambda_{cN}(T)$ which proves the proposition 1.
\paragraph*{}Note that imposing the uniform convergence of $C(x,s,t)$ to
$C(0,s,t)$ is not a very restrictive condition. As far as we know, it
seems to be fulfilled by any nonpathological stochastic field $S$ of
physical interest.
%
%
\section{Equality of $\lambda_{cN}(T)$ and $\overline{\lambda}_{cN}(T)$
for a class of $S$}\label{sec4}
{}For a large class of Gaussian fields $S$ it is possible to prove that
diffusion has no effect on the onset of the divergence of $\langle
{\cal E}(x,T)^N\rangle$, i.e. $\lambda_{CN}(T) =
\overline{\lambda}_{cN}(T)$.
\paragraph*{}
\bigskip
\noindent {\bf Proposition 2}:
Assume that for every $T>0$ one has $\lim_{x\rightarrow 0}\sup_{s,t\in
[0,T]} \vert C(x,s,t)-C(0,s,t)\vert =0$, and that $\vert C(x,s,t)\vert\le
C(0,s,t)$ for every $x\in {\mathbb R}^d$ and $s,t\in [0,T]$. Then
$\lambda_{cN}(T)=\overline{\lambda}_{cN}(T)$.
\paragraph*{}
\bigskip
The proof of this proposition is as follows: By the uniform convergence
condition on $C(x,s,t)$ and Proposition 1 one already has
$\lambda_{cN}(T)\le \overline{\lambda}_{cN}(T)$. It remains to show
that $\overline{\lambda}_{cN}(T)\le \lambda_{cN}(T)$. Let $\mu_1$ be
the largest eigenvalue of the covariance operator $\hat{t}_{C_0}$. Let
$\phi_1({\bf t})$ be a principal (normalized) eigenvector for the
covariance operator $\hat{T}_C$. One has
$$
\mu_1^{x({\bf t})}=(\phi_1,\hat{T}_C\phi_1)\le
(\vert\phi_1\vert,\vert \hat{T}_C\phi_1\vert)\le
(\vert\phi_1\vert,\hat{T}_{C_0}\vert\phi_1\vert)\le N\mu_1,
$$
where the second inequality follows from the condition $\vert
C(x,s,t)\vert\le C(0,s,t)$. Suppose now
$\lambda<\overline{\lambda}_{cN}(T)$, i.e. $\lambda <(2N\mu_1)^{-1}$.
Then $\lambda <(2\mu_1^{x({\bf t})})^{-1}$ and by Lemma 1
$$
\left\langle\exp
\left\lbrack\lambda\int S({\bf t})^2d{\bf t}\right\rbrack
\right\rangle
\le\exp\left\lbrack\frac{N\lambda \int_0^TC(0,t,t)\,
dt}{1-2\lambda\mu_1^{x({\bf t})}}\right\rbrack
\le\exp\left(\frac{N\lambda \int_0^TC(0,t,t)\,
dt}{1-2N\lambda\mu_1}\right).
$$
Since this inequality is uniform over all Brownian paths, we finally have
$$
\langle {\cal E}(0,T)^N\rangle
\le\exp\left(\frac{N\lambda \int_0^TC(0,t,t)\,
dt}{1-2N\lambda\mu_1}\right)<+\infty,
$$
and therefore $\lambda<\lambda_{cN}(T)$, which proves the proposition 2.
\paragraph*{}This result shows that for Gaussian fields $S$ fulfilling
the not so restrictive conditions of Proposition 2, it is sufficient to
solve the diffusion-free problem to determine the onset of the
divergence of $\langle {\cal E}(x,T)^N\rangle$. It is therefore
interesting to show how such fields can be actually obtained. To this
end, the remaining of this section will be devoted to explicitely
construct two typical examples of stochastic fields $S$ which fulfill
the conditions of Proposition 2.
\subsection{an example of nonstationary $S$}\label{sec4.1}
The first example is the diffusive counterpart of the Gaussian field
defined by Eq.\ (\ref{eq1.2}). Let $S(x,t)$ be the solution to
\begin{equation}\label{eq4.1}
\left\lbrace\begin{array}{l}
\partial_tS(x,t)-\frac{1}{2}\Delta S(x,t)=0,\\
t\in [0,T],\ x\in {\mathbb R}^d,\ {\rm and}\
S(x,0)={\cal S}(x),
\end{array}\right.
\end{equation}
where ${\cal S}(x)$ is a real homogeneous Gaussian field defined by
\begin{equation}\label{eq4.2}
\begin{array}{l}
\langle {\cal S}(x)\rangle =0,\\
\langle {\cal S}(x){\cal S}(x')\rangle ={\cal C}(x-x'),
\end{array}
\end{equation}
with ${\cal C}(x)$ a given\ \cite{note1} function of $x$ normalized such
that ${\cal C}(0)\equiv\langle S(x,0)^2\rangle=1$. One has
\begin{equation}\label{eq4.3}
S(x,t)=\int {\cal S}(k){\rm e}^{ikx-\frac{1}{2}k^2t}d^dk,
\end{equation}
where ${\cal S}(k)$ is the {}Fourier transform of ${\cal S}(x)$, and from
Eqs.\ (\ref{eq4.2}) and\ (\ref{eq4.3}) it follows that $S(x,t)$ is a real
homogeneous nonstationary Gaussian field with
\begin{equation}\label{eq4.4}
\begin{array}{l}
\langle S(x,t)\rangle =0,\\
\langle S(x,t)S(x',t')\rangle =\int {\cal C}(k)
{\rm e}^{ik(x-x')-\frac{1}{2}k^2(t+t')}d^dk,
\end{array}
\end{equation}
where ${\cal C}(k)$ is the {}Fourier transform of ${\cal C}(x)$. Since
${\cal C}(k)$ is real and positive\ \cite{note1}, one has
\begin{eqnarray*}
&&\vert C(x,s,t)\vert\equiv\vert\langle S(x,s)S(0,t)\rangle\vert\\
&&=\left\vert\int {\cal C}(k)
{\rm e}^{ikx-\frac{1}{2}k^2(s+t)}d^dk\right\vert\\
&&\le\int {\cal C}(k)
{\rm e}^{-\frac{1}{2}k^2(s+t)}d^dk=C(0,s,t),
\end{eqnarray*}
for every $x\in {\mathbb R}^d$ and $s,t\in [0,T]$, so $S(x,t)$ fulfills
the conditions of Proposition 2.
\subsection{an example of stationary $S$}\label{sec4.2}
The second example is provided by a modified version of Eq.\
(\ref{eq4.1}) obtained by adding a source term \`a la Langevin on its
right-hand side. Namely, let $S(x,t)$ be the solution to
\begin{equation}\label{eq4.5}
\left\lbrace\begin{array}{l}
\partial_tS(x,t)-\frac{1}{2}\Delta S(x,t)=L(x,t),\\
t\in ]-\infty ,T],\ x\in {\mathbb R}^d,\ {\rm and}\
S(x,-\infty)=0,
\end{array}\right.
\end{equation}
where the Langevin source term $L(x,t)$ is a homogeneous Gaussian white
noise defined by
\begin{equation}\label{eq4.6}
\begin{array}{l}
\langle L(x,t)\rangle =0,\\
\langle L(x,t)L(x',t')\rangle =-\delta(t-t')\Delta_x{\cal C}(x-x'),
\end{array}
\end{equation}
with ${\cal C}(x)$ a given\ \cite{note1} function of $x$ normalized such
that ${\cal C}(0)=1$. The solution to Eq.\ (\ref{eq4.5}) reads
\begin{equation}\label{eq4.7}
S(x,t)=\int d^dk\left\lbrack
{\rm e}^{ikx}\int_{-\infty}^t{\rm e}^{-\frac{1}{2}k^2(t-s)}L(k,s)\, ds
\right\rbrack,
\end{equation}
where $L(k,t)$ is the {}Fourier transform of $L(x,t)$. From Eqs.\
(\ref{eq4.6}) and\ (\ref{eq4.7}) it can be shown that $S(x,t)$ is a
real homogeneous stationary Gaussian field with
\begin{equation}\label{eq4.8}
\begin{array}{l}
\langle S(x,t)\rangle =0,\\
\langle S(x,t)S(x',t')\rangle =\int {\cal C}(k)
{\rm e}^{ik(x-x')-\frac{1}{2}k^2\vert t-t'\vert}d^dk,
\end{array}
\end{equation}
where ${\cal C}(k)$ is the {}Fourier transform of ${\cal C}(x)$. As
previously, since ${\cal C}(k)$ is real and positive\ \cite{note1}, one
has
\begin{eqnarray*}
&&\vert C(x,s,t)\vert\equiv\vert\langle S(x,s)S(0,t)\rangle\vert\\
&&=\left\vert\int {\cal C}(k)
{\rm e}^{ikx-\frac{1}{2}k^2\vert s-t\vert}d^dk\right\vert\\
&&\le\int {\cal C}(k)
{\rm e}^{-\frac{1}{2}k^2\vert s-t\vert}d^dk=C(0,s,t),
\end{eqnarray*}
for every $x\in {\mathbb R}^d$ and $s,t\in [0,T]$, and so $S(x,t)$
fulfills the conditions of Proposition 2.
\paragraph*{} More generally, it can be checked that any
real homogeneous Gaussian field $S(x,t)$ defined by
$$
\begin{array}{l}
\langle S(x,t)\rangle =0,\\
\langle S(x,t)S(x',t')\rangle =
\int {\cal C}(k,t,t'){\rm e}^{ik(x-x')}d^dk,
\end{array}
$$
where ${\cal C}(k,t,t')$ is real and positive, fulfills the
conditions of Proposition 2.
%
%
\section{Explicit solution of the diffusion-free problem for a class of
$S$}\label{sec5}
In this section we show that an explicit computation of the
diffusion-free amplification factor $\langle \exp (N\lambda\int_0^T
S(0,t)^2dt)\rangle$ can be achieved if $S(0,t)$ is a linear functional
of a Gauss-Markov process. Note that determining
$\overline{\lambda}_{cN}(T)$ amounts to finding the largest eigenvalue
of the covariance operator $\hat{t}_{C_0}$, which in
principle can always be achieved, at least numerically. As shown above,
$\overline{\lambda}_{cN}(T) \ge\lambda_{cN}(T)$ with equality holding
when Proposition 2 is applicable. Since
$\overline{\lambda}_{cN}(T)=N^{-1} \overline{\lambda}_{c1}(T)$ in the
diffusion free case, we will take $N=1$ in the remaining of this
section without loss of generality.
\subsection{Solution of the diffusion-free problem using the {}Feynman-Kac
formula}\label{sec5.1}
We consider the case where the Gaussian process $S(0,t)$ can be written
as
\begin{equation}\label{eq5.1}
S(0,t)=\langle c,Y(t)\rangle,
\end{equation}
where $\langle x,y\rangle\equiv x^\dag y=\sum_ix_iy_i$, $c$ is a
given n-dimensional vector, and $Y(t)$ is a n-dimensional Gauss-Markov
process defined as the solution to the linear stochastic differential
equation
\begin{equation}\label{eq5.2}
\left\lbrace\begin{array}{l}
dY(t)+AY(t)dt=GdB(t),\\
Y(0)\ {\rm Gaussian\ with\ }\langle Y(0)\rangle =0.
\end{array}\right.
\end{equation}
Here, $A$ and $G$ are constant $n\times n$ matrices, and $B(t)$ is a
$n$-dimensional Brownian motion. {}From Eqs.\ (\ref{eq5.1}) and\
(\ref{eq5.2}), it follows that one can write the diffusion-free
amplification factor as a {}Feynman-Kac formula
\begin{equation}\label{eq5.3}
\left\langle {\rm e}^{\lambda\int_0^TS(0,t)^2\, dt}\right\rangle =
\left\langle {\rm e}^{\lambda\int_0^T\langle Y(t),CY(t)\rangle\,
dt}\right\rangle
=\int v(y,T)\, d^ny,
\end{equation}
where $C$ denotes the symmetrical $n\times n$ matrix $c\otimes c$,
and $v(y,t)$ is the solution to the parabolic equation
\begin{equation}\label{eq5.4}
\left\lbrace\begin{array}{l}
\frac{\partial v}{\partial t}=
({\rm Tr}A+\lambda \langle y,Cy\rangle)v+
\langle Ay,\nabla\rangle v+
\frac{1}{2}\langle G^\dag\nabla ,G^\dag\nabla\rangle v,\\
v(y,0)=\left(\frac{1}{2\pi}\right)^{n/2}
\frac{1}{\sqrt{\vert U\vert}}
\exp\left\lbrack -\frac{1}{2}
\langle y,U^{-1}y\rangle
\right\rbrack,
\end{array}\right.
\end{equation}
with $U={\rm Cov}[Y(0),Y(0)]$. The solution to Eq.\ (\ref{eq5.4}) has
the form
\begin{equation}\label{eq5.5}
v(y,t)=\left(\frac{1}{2\pi}\right)^{n/2}
\frac{1}{\sqrt{\vert K(t)\vert}}
\exp\left\lbrack -\frac{1}{2}
\langle y,K(t)^{-1}y\rangle
+\lambda\int_0^t{\rm Tr}CK(s)\, ds\right\rbrack,
\end{equation}
where $K(t)$ is a symmetrical $n\times n$ matrix which is the
solution to
\begin{equation}\label{eq5.6}
\left\lbrace\begin{array}{l}
\frac{dK(t)}{dt}=
GG^\dag-[AK(t)+K(t)A^\dag]+2\lambda K(t)CK(t),\\
K(0)=U.
\end{array}\right.
\end{equation}
Thus, from Eqs.\ (\ref{eq5.3}) and\ (\ref{eq5.5}) one has
\begin{equation}\label{eq5.7}
\left\langle {\rm e}^{\lambda\int_0^TS(0,t)^2\, dt}\right\rangle =
{\rm e}^{\lambda\int_0^T{\rm Tr}CK(t)\, dt}.
\end{equation}
with $K(t)$ given by the Riccati equation\ (\ref{eq5.6}).
\paragraph*{}The solution to Eq.\ (\ref{eq5.6}) is known to explode in
finite time for large enough $\lambda$. {}For $n=1$, in which case
$S(0,t)$ is itself Markovian, Eq.\ (\ref{eq5.6}) is solved
straightforwardly (see Sec.\ \ref{sec5.2}). {}For $n\ge 2$, the solution
to Eq.\ (\ref{eq5.6}) can be obtained by the so-called Hamiltonian
method: we define the $2n\times 2n$ matrix
$$
H = \left(\begin{array}{cc}
A^\dag & -2\lambda C \\ GG^\dag & -A
\end{array}
\right)
$$
and solve the linear differential equation
\begin{equation}\label{eq5.8}
\frac{d}{dt}\left\lbrack\begin{array}{c}
Q(t) \\ P(t)
\end{array}\right\rbrack
= H\left\lbrack\begin{array}{c}
Q(t) \\ P(t)
\end{array}\right\rbrack,
\end{equation}
with the initial condition
$$
\left\lbrack\begin{array}{c}
Q(0) \\ P(0)
\end{array}\right\rbrack
= \left\lbrack\begin{array}{c}
I \\ U
\end{array}\right\rbrack.
$$
The solution $K(t)$ to the Riccati equation\ (\ref{eq5.6}) is
easily checked to be given by
\begin{equation}\label{eq5.9}
K(t)=P(t)Q(t)^{-1},
\end{equation}
which explodes if and only if $Q(t)$ becomes singular\ \cite{Crouch}.
Since Eq.\ (\ref{eq5.8}) is a linear equation, it can in principle be
solved by a symbolic computation program.
\subsection{Application to the ${\bf n=1}$ case}
\label{sec5.2}
As an example, let us consider the simplest
case $n=1$ with $C(0,t,t')={\rm e}^{-\vert t-t'\vert}$. In this limit,
the diffusion-free amplification factor reads
\begin{equation}\label{eq5.10}
\left\langle {\rm e}^{\lambda\int_0^TS(0,t)^2\, dt}\right\rangle =
\left\langle {\rm e}^{\lambda\int_0^T Y(t)^2\,
dt}\right\rangle
={\rm e}^{\lambda\int_0^TK(t)\, dt},
\end{equation}
where $Y(t)$ is the Ornstein-Uhlenbeck process
\begin{equation}\label{eq5.11}
\left\lbrace\begin{array}{l}
dY(t)+Y(t)dt=\sqrt{2}dB(t),\\
\langle Y(0)\rangle =0,\ \langle Y(0)^2\rangle =1,
\end{array}\right.
\end{equation}
and $K(t)$ is the solution to the Riccati equation
\begin{equation}\label{eq5.12}
\left\lbrace\begin{array}{l}
\frac{1}{2}\frac{dK(t)}{dt}=
1-K(t)+\lambda K(t)^2,\\
K(0)=1.
\end{array}\right.
\end{equation}
Equation\ (\ref{eq5.12}) can be easily solved by means of the
substitution $2\lambda K(t)=-d\log u(t)/dt$. Inserting the
result into Eq.\ (\ref{eq5.10}), one obtains
\begin{equation}\label{eq5.13}
\left\langle {\rm e}^{\lambda\int_0^TS(0,t)^2\, dt}\right\rangle =
\frac{{\rm e}^{T/2}}{\sqrt{\cosh(\alpha T)+\alpha^{-1}(1-2\lambda)
\sinh(\alpha T)}}, \quad \lambda < 1/4,
\end{equation}
\begin{equation}\label{eq5.14}
\left\langle {\rm e}^{\lambda\int_0^TS(0,t)^2\, dt}\right\rangle =
\frac{{\rm e}^{T/2}}{\sqrt{1+T/2}}, \quad \lambda = 1/4,
\end{equation}
\begin{equation}\label{eq5.15}
\left\langle {\rm e}^{\lambda\int_0^TS(0,t)^2\, dt}\right\rangle =
\frac{{\rm e}^{T/2}}{\sqrt{\cos(\alpha T)+\alpha^{-1}(1-2\lambda)
\sin(\alpha T)}}, \quad \lambda > 1/4,
\end{equation}
where $\alpha =\vert 1-4\lambda\vert^{1/2}$. It
can be seen from Eq.\ (\ref{eq5.15}) that, for $\lambda > 1/4$,
$\langle \exp (\lambda\int_0^T S(0,t)^2dt)\rangle$ diverges as $T$
tends (from below) to the critical time $T_c(\lambda)$ given by
\begin{equation}\label{eq5.16}
T_c(\lambda)=\frac{1}{\sqrt{4\lambda -1}}
\tan^{-1}\left(\frac{\sqrt{4\lambda -1}}{2\lambda -1}\right),
\end{equation}
where the determination of $\tan^{-1}$ is such that $0<\tan^{-1}\le\pi$.
Inverting Eq.\ (\ref{eq5.16}) and using
$\overline{\lambda}_{cN}(T)=N^{-1} \overline{\lambda}_{c1}(T)$ gives
the diffusion-free critical coupling $\overline{\lambda}_{cN}(T)$ in
the cases where $C(0,t,t')={\rm e}^{-\vert t-t'\vert}$.
%
%
\section{Dependence of the critical coupling on space dimensionality}
\label{sec6}
In this section we study the dependence of $\lambda_{cN}(T)$ on the
space dimensionality $D$. We will restrict ourselves to the cases where
the correlation function $C$ can be written out as
\begin{equation}\label{eq6.1}
C_D(x,t,t')=C_d(x_{||},t,t')C_{D-d}(x_\bot,t,t'),
\end{equation}
with $C_D$, $C_d$ and $C_{D-d}$ continuous, symmetric, and positive
definite, and where $x_{||}$ is the projection of $x$ onto a
$d$-dimensional subspace ($d<D$) and $x_\bot=x-x_{||}$. In the
following, a correlation function of this type will be called a
factorable correlation function. It is worth noting that such a
correlation function can be very easily obtained, e.g. when the
Gaussian field $S$ is defined by either Eq.\ (\ref{eq4.4}) or Eq.\
(\ref{eq4.8}) in the cases where ${\cal C}(k)$ is factorable as ${\cal
C}(k)={\cal C}_d(k_{||}){\cal C}_{D-d}(k_\bot)$.
\paragraph*{}We prove that as $\lambda$ increases, the divergence of
$\langle {\cal E}(x,T)^N\rangle$ obtained in the original
$D$-dimensional problem cannot occur before the one obtained in the
projected $d$-dimensional problem whenever $0\le C_{D-d}(0,t,t)\le 1$.
Since many stochastic fields $S$ of physical interest, e.g. in optics,
do have a factorable correlation function, we expect this result to be
useful for the comparison of two-dimensional numerical simulations with
experiments and three-dimensional numerical simulations. Before
expressing this result in a more rigorous way, we begin with two
technical lemmas that will be needed in the following.
\paragraph*{}
\bigskip
\noindent {\bf Lemma 3}:
Consider a $D$-dimensional problem and let $\mu_1^{x({\bf t})}$ be the
largest eigenvalue of the covariance operator $\hat{T}_{C_D}$ and N
given continuous paths $x({\bf t})$. Then $\lambda_{cN}(T,D)=
[2\sup_{x({\bf t})} \mu_1^{x({\bf t})}]^{-1}$.
\paragraph*{}
\bigskip
This lemma can be proven straightforwardly by successively
considering the inequalities $\lambda>[2\sup_{x({\bf t})}\mu_1^{x({\bf
t})}]^{-1}$ and $\lambda<[2\sup_{x({\bf t})}\mu_1^{x({\bf t})}]^{-1}$,
and by following the same lines of reasoning as for the proofs of
Propositions 1 and 2 respectively, where one replaces the N paths
$x({\bf t})=0$ corresponding to $\overline{\lambda}_{cN}(T)=
[2\mu_1^{x({\bf t})=0}]^{-1}$ by N paths maximizing $\mu_1^{x({\bf
t})}$\ \cite{note}.
\paragraph*{}
\bigskip
\noindent {\bf Lemma 4}:
Let $K_0({\bf s},{\bf t})$, $K_1({\bf s},{\bf t})$, and $K_2({\bf
s},{\bf t})$ be three symmetric kernels such that: (i) $K_0({\bf
s},{\bf t})=K_1({\bf s},{\bf t})K_2({\bf s},{\bf t})$; (ii) $K_2$ is a
positive definite continuous symmetric kernel; (iii) $0\le K_2({\bf
t},{\bf t})<1$ and the largest eigenvalue of $K_1$ is positive, or
$K_2({\bf t},{\bf t})=1$ and no condition on the sign of the largest
eigenvalue of $K_1$. Then $\mu_1(K_0)\le \mu_1(K_1)$, where
$\mu_1(K_\alpha)$ denotes the largest eigenvalue of $K_\alpha$.
\paragraph*{}
\bigskip
The proof of this lemma is as follows: since $K_2$ is a
positive definite continuous symmetric kernel, Mercer's theorem holds\
\cite{CH} and this kernel admits the expansion
\begin{equation}\label{eq6.2}
K_2({\bf s},{\bf t})=\sum_i a_if_i({\bf s})f_i({\bf t}),
\end{equation}
where $a_i\ge 0$ and $f_i({\bf t})$ respectively denote the $i^{th}$
eigenvalue of the operator $\hat{T}_{K_2}$ and the associated
normalized eigenfunction. Let $\phi_1({\bf t})$ be a principal
(normalized) eigenfunction of the operator $\hat{T}_{K_0}$ and
$\mu_1(K_0)$ the corresponding largest eigenvalue. {}From the condition
(i) and Eq.\ (\ref{eq6.2}), one has
\begin{equation}\label{eq6.3}
\mu_1(K_0)=(\phi_1,\hat{T}_{K_0}\phi_1)=
\sum_i a_i(f_i\phi_1,\hat{T}_{K_1}f_i\phi_1)=
\sum_i a_iM_i(\eta_i,\hat{T}_{K_1}\eta_i),
\end{equation}
where $M_i$ and $\eta_i({\bf t})$ are given by
$$
M_i=(f_i\phi_1,f_i\phi_1),
$$
and
$$
\eta_i({\bf t}) =M_i^{-1/2}f_i({\bf t})\phi_1({\bf t}),
$$
such that $(\eta_i,\eta_i)=1$. By the definition of
$\mu_1(K_1)$ and from $K_2({\bf t},{\bf
t})\le 1$, condition (iii), one has respectively
\begin{equation}\label{eq6.4}
\mu_1(K_1)\ge (\eta_i,\hat{T}_{K_1}\eta_i),
\end{equation}
and
\begin{eqnarray}\label{eq6.5}
&&\sum_i a_iM_i=\int\left\lbrack\sum_i a_i f_i({\bf
t})^2\right\rbrack\phi_1({\bf t})^2d{\bf t}\nonumber\\
&&=\int K_2({\bf
t},{\bf t})\phi_1({\bf t})^2d{\bf t}\le
\int\phi_1({\bf t})^2d{\bf t}=1.
\end{eqnarray}
So, from Eqs.\ (\ref{eq6.3}), (\ref{eq6.4}), (\ref{eq6.5}) and the
condition (iii), it follows that $\mu_1(K_0)\le \mu_1(K_1)$, which
proves Lemma 4.
\paragraph*{}We can now proceed to rigorously express and prove the
result stated at the beginning of this section. Let $\lambda_{cN}(T,D)$
be the critical coupling associated with a $D$-dimensional problem in
which the correlation function of the Gaussian field $S$ is given by
$C_D$. One has the following proposition:
\paragraph*{}
\bigskip
\noindent {\bf Proposition 3}:
for every $T>0$, if $C_D(x,t,t')$ is a factorable correlation function
such that $0\le C_{D-d}(0,t,t)\le 1$ for $0\le t\le T$, then
$\lambda_{cN}(T,D)\ge\lambda_{cN}(T,d)$.
\paragraph*{}
\bigskip
The proof of this proposition is straightforward. By the definition of a
factorable correlation function one has $C_D({\bf s},{\bf t})=C_d({\bf
s},{\bf t})C_{D-d}({\bf s},{\bf t})$, where both $C_d({\bf s},{\bf t})$
and $C_{D-d}({\bf s},{\bf t})$ are continuous, symmetric, and positive
definite. Since $C_{D-d}({\bf t},{\bf t})\equiv C_{D-d}(0,t,t)$ and
$0\le C_{D-d}(0,t,t)\le 1$ by assumption, one can apply the lemma 4
with $K_0=C_D$, $K_1=C_d$, and $K_2=C_{D-d}$. It follows immediately
that $\mu_1^{x({\bf t})}\le\tilde\mu_1^{x_{||}({\bf t})}$, where
$\tilde\mu_1^{x_{||}({\bf t})}$ denotes the largest eigenvalue of the
operator $\hat{T}_{C_d}$. Let $x_{\rm max}({\bf t})$ be N paths
maximizing $\mu_1^{x({\bf t})}$\ \cite{note}. One has $\sup_{x({\bf
t})}\mu_1^{x({\bf t})}= \mu_1^{x_{\rm max}({\bf
t})}\le\tilde\mu_1^{x_{{\rm max}||}({\bf t})}$, from which it follows
that $\sup_{x({\bf t})}\mu_1^{x({\bf t})}\le \sup_{x_{||}({\bf
t})}\tilde\mu_1^{x_{||}({\bf t})}$ and, by Lemma 3,
$\lambda_{cN}(T,D)\ge\lambda_{cN}(T,d)$, which proves the proposition
3.
%
%
\section{Summary and perspectives}\label{sec7}
In this paper, we have studied the effects of diffusion on the
divergence of the moments of the solution to a linear amplifier driven
by the square of a Gaussian field. We first proved that the divergence
yielded by a diffusion-free calculation cannot occur at a smaller
coupling constant than the one obtained from the full calculation (i.e.
with diffusion). Then we have shown that, in the case where the
absolute value of the (uniformly continuous) pump field correlation
function is bounded from above by its one-point value, there is no
diffusion effect on the onset of the divergence which is therefore
given by a diffusion-free calculation. In this context, we have solved
the diffusion-free problem explicitly when the pump field is a linear
functional of a Gauss-Markov process. {}Finally, we have studied the
dependence of the critical coupling on the space dimensionality in the
case of a factorable correlation function. In particular, we have
proved that the divergence obtained in a $D$-dimensional problem cannot
occur at a smaller coupling constant than the one obtained in the
projected $d$-dimensional problem ($d<D$).
\paragraph*{}As mentioned in the introduction, we would like to extend
our results for the diffusion-amplification model\ (\ref{eq2.1}) to the
more difficult diffraction-amplification problem\ (\ref{eq1.1}).
According to Eq.\ (\ref{eq2.1}), the results obtained in this paper
also apply, beside some minor technical modifications, if the pump
field is a {\it complex} Gaussian field as in Eq.\ (\ref{eq1.1}). The
remaining difficulty in extending our results to Eq.\ (\ref{eq1.1})
lies in controlling the complex {}Feynman path-integral, compared to that
of the Feynman-Kac formula for the diffusive case. Expressing ${\cal E}
(x,t)$ as a {}Feynman path-integral and averaging over the realizations
of $S$, one cannot {\it a priori} exclude the possibility that
destructive interference effects between different path contributions
make the sum of the divergent contributions finite. Thus one cannot
deduce the divergence of the moments of ${\cal E}(x,L)$ from that of
the amplification along paths arriving at the point $(x,L)$. It is
however not unreasonable to expect that Propositions 1, 2, and 3 also
apply to the diffraction-amplification problem\ (\ref{eq1.1}). Proving
this conjecture is another matter and is the subject of a future work.
Note that in the case of Proposition 2, the on-axis correlation
function of the pump field must be real and positive, which is quite
restrictive if the pump field is complex. {}From a practical point of
view (e.g. in optics), it would therefore be very interesting to find
out whether there exists an enlarged version of this proposition
applying to complex on-axis correlation functions as well.
%
%
\section{Acknowledgments}\label{sec8}
We thank Harvey Rose for introducing us to this problem and for
providing many valuable insights. The work of J.L.L. was supported in
part by AFOSR Grant F49620-98-1-0207 and NSF Grant DMR-9813268.  A.A.,
P.D.P. and J.L.L. acknowledge the hospitality of the IHES at
Bures-sur-Yvette, {}France, where part of this work was done.
%
%

%

\begin{references}
\bibitem{note1} Since ${\cal C}(x)$ is a covariance, it must be chosen in
such a way that ${\cal C}(x-x')$ is a positive definite kernel, i.e.
its Fourier transform must be real, even, and positive. Note that
$C(x,z)$ with $z\in [-L,0[$ can be obtained straightforwardly from\
(2) and the Hermitian symmetry $C(x,-z)=C(-x,z)^\ast$.
\bibitem{note} In the cases where there is no such a set of paths, one
should consider N paths that realize the supremum up to a arbitrarily
small constant.
<%
\end{references}
\end{document}